\newcommand{\simgt}{\lower.5ex\hbox{$\; \buildrel > \over \sim \;$}}
\newcommand{\simlt}{\lower.5ex\hbox{$\; \buildrel < \over \sim \;$}}
\newcommand{\citet}[1] {\cite{#1}}
\newcommand{\citep}[1] {(\cite{#1})}
\newcommand{\bm}[1]{\mbox{\boldmath$#1$}}

\newcommand{\mnras}{Mon. Not. R. Astron. Soc.}
\documentstyle[prl,aps,epsf]{revtex}
\begin{document}
\twocolumn[\hsize\textwidth\columnwidth\hsize\csname@twocolumnfalse\endcsname

\title{Number Count of Peaks in the CMB Map}
\author{\sc Toshifumi Futamase, and Masahiro Takada}
\address{%
Astronomical Institute, Graduate School of Science,
Tohoku University, Sendai 980-8578, Japan;
tof, takada@astr.tohoku.ac.jp}

\maketitle
\begin{abstract}
We investigate  the dependence of cosmological parameters on 
the number count of peaks (local maxima and minima) 
in the cosmic microwave background (CMB) sky.  
The peak statistics contains the whole information of 
acoustic oscillations in the angular power spectrum $C_l$
over $l$-space and thus it can place complementary
constraints on the cosmological parameters
to those obtained from measurements of $C_l$.  Based on
the instrumental specifications of Planck, 
we find that the number count of peaks can provide new
constraints on the combination of the matter density $\Omega_{\rm m}$
and the Hubble parameter $h$  approximately scaled as
$\Omega_{\rm m} h^{-4.9}$ for a flat $\Lambda$CDM model
with $\Omega_{m}=0.3$ and $h=0.7$. Therefore, we suggest
that combining it with the constraints from $C_l$ scaled as
 $\Omega_{m}h^{3.8}$ (or commonly $\Omega_{\rm m}h^2$) 
can potentially determine $\Omega_{\rm m}$ or equivalently 
 solve the cosmic degeneracy by the CMB data alone. 
\end{abstract}
] 

With the data from the BOOMERanG \cite{boom} and MAXIMA
\cite{Maxima} experiments, the cosmic microwave background (CMB)
is dramatically improving our knowledge of cosmological 
parameters \cite{Lange,TZ,hu}. 
Those data have revealed that the measured angular
power spectrum $C_l$ is fairly consistent with that predicted by
the adiabatic inflationary models. Then, the position of the first
acoustic peak in $C_l$ is very sensitive to the spatial curvature or
equivalently the total density of the universe \cite{JKKS}, and 
the measured position has favored a flat geometry of the universe.
Furthermore, if we adopt the simplest flat inflationary models with
purely scalar scale-invariant fluctuations,
the position can place constraints on the combination of the
matter energy density $\Omega_{\rm m}$ and the Hubble parameter $h$
with the dependence of $\Omega_{\rm m}h^2$ \cite{TZ} or
$\Omega_{\rm m}h^{3.8}$ in more detailed analysis \cite{hu}. 
However, it still remains difficult to accurately
determine $\Omega_{\rm m}$ only from the measurements of $C_l$
(so-called the {\em cosmic degeneracy} \cite{BET}). 
The inflationary scenarios also predict that the primordial
fluctuations are Gaussian \cite{Guth}, and in this case 
the peak statistics could provide an additional information
about statistical properties of the
distributions of peaks (local maxima or minima) in the CMB map
based on the Gaussian theory \cite{BE,HS,Takada}.
One of the most straightforward statistical measures in the peak statistics
is then the mean number density or the number count of peaks.
We expect that the number count of peaks
can place complementary
constraints on the cosmological parameters to those provided by
the measurements of $C_l$ because the number count depends on
spectral parameters obtained from the {\em integration} of some
weighted $C_l$ over $l$ space. 


Moreover, there are some advantages of using the peak statistics.  
Since the number count of peaks is given as
a function of a certain threshold in units of the rms
of the temperature fluctuations itself, we expect that the
peak statistics is more robust against the systematic observational
errors of CMB anisotropies. 

In this {\it Letter}, therefore, we present theoretical predictions of peak number 
count in the CMB maps taking into account instrumental effects of 
beam size and detector noise. We then focus our investigations
on a problem how the number count for the observed sky
coverage can place complementary constrains on $\Omega_{m}$-$h$
plane by fixing other cosmological parameters.
This is motivated by our expectation
that the number count can potentially break the
cosmic degeneracy. 

Under the Gaussian assumption for the primordial fluctuations,
statistical properties of any primary CMB field can be exactly 
computed once $C_l$ is given. It is then convenient to introduce 
the spectral parameters \cite{BE}:
\begin{equation}
\sigma_{\rm n}^2=\int\!\!\frac{ldl}{2\pi}C_l l^{2{\rm n}}, 
\hspace{2em}\gamma\equiv\frac{\sigma_1^2}{\sigma_0\sigma_2},
\hspace{2em}\theta_\ast\equiv\sqrt{2}\frac{\sigma_1}{\sigma_2}.
\label{eqn:spec}
\end{equation}
Note that $\sigma_0$, $\sigma_1$, and $\sigma_2$ 
represent the rms values of temperature fluctuations field
($=\Delta\equiv\delta T/T_{\rm CMB}$), its gradient and second spatial
derivative fields, respectively, and $\sigma_0$ is of the order of $\sigma_0\sim
5\times 10^{-5}$ in the cosmological models we consider.
The parameter $\theta_\ast$ gives the characteristic 
curvature scale of the temperature fluctuations field and can be
estimated as $\theta_\ast\sim 5'$. Throughout this {\it Letter}
we employ the small angle approximation \cite{BE}
where we use the Fourier analysis in the two-dimensional flat space
instead of in the spherical space. 
It will be a good approximation for our arguments
because modes of $C_l$ at $l\simgt100$ produce dominant
contributions to $\sigma_1$ and $\sigma_2$ that are
main parameters to control how many peaks (local maixima or minima)
are generated in the observed CMB sky \cite{Takada}.
We can then use the following analytic expression \cite{BE}
for the differential number density of local maxima ({\em hotspots})
of height in the range of $\nu(\equiv\Delta_{\rm
peak}/\sigma_0)$ and $\nu+d\nu$:
\begin{equation}
dn_{\rm hotspots}(\nu)
=\frac{1}{(2\pi)^{3/2}\theta_\ast^2}\exp\left[-\frac{\nu^2}{2}\right]
G(\gamma,\gamma\nu)d\nu
\label{eqn:dn}
\end{equation}
where
\begin{eqnarray}
G(\gamma,x_\ast)&\equiv&
(x_\ast^2-\gamma^2)\left\{1-\frac{1}{2}{\rm erfc}\left[
 \frac{x_\ast}{\sqrt{2(1-\gamma^2)}}\right]\right\}
\nonumber\\
&&\hspace{-5em}+x_\ast(1-\gamma^2)
\frac{\exp\{-x_\ast^2/
[2(1-\gamma^2)]\}}{\sqrt{2\pi(1-\gamma^2)}}\!+\!\frac{
\exp[-x_\ast^2/(3-2\gamma^2)]}{(3-2\gamma^2)^{1/2}}
\nonumber\\
&&\times\left\{1-\frac{1}{2}{\rm erfc}\left[\frac{x_\ast}{\sqrt{2
(1-\gamma^2)(3-2\gamma^2)}}\right]
 \right\}.
 \label{eqn:diffnum}
\end{eqnarray}
The function ${\rm erfc}(x)$ denotes the Gaussian error
function defined by ${\rm erfc}(x)=2/\sqrt{\pi}
\int^{\infty}_x\!\!dt \exp[-t^2]$.
Equation (\ref{eqn:dn}) clearly shows that the number
density of peaks increases with a decrease of $\theta_\ast$
and $\gamma$ affects it through the dependence on the function $G$.
As discussed later, $\theta_\ast$ is very sensitive to the position
of acoustic peaks of $C_l$, which largely
depends on the cosmological parameters \cite{hu},
because the sound horizon at decoupling is the characteristic
wavelength of temperature fluctuations.
The mean number density of hotspots 
of height above $\nu_{\rm s}$ can be obtained by integrating
equation (\ref{eqn:dn}) over $\nu>\nu_{\rm s}$.
In the Gaussian field, the number density of 
local minima ({\em coldspots}) of height below $-\nu_{\rm s}$ is 
also given by $n_{\rm coldspots}(<-\nu_{\rm s})
=n_{\rm hotspots}(>\nu_{\rm s})$ due to the symmetric
nature.   The total mean number density of such 
maxima and minima per unit solid angle is defined by
$n_{\rm peak}(\nu_{\rm s})=
n_{\rm hotspots}(>\nu_{\rm s})+n_{\rm coldspots}(<-\nu_{\rm s})$. 
Therefore, if the sky coverage is $f_{\rm sky}(\le1)$,
the number count of such peaks in the observed CMB sky can be predicted
as
\begin{equation}
N_{\rm peak}(\nu_{\rm s})=\Omega_{\rm sky}n_{\rm peak}(\nu_{\rm s})
 =4\pi f_{\rm sky }n_{\rm peak}(\nu_{\rm s}). 
\end{equation}
This is a basic equation used in the following discussions. 

For a practical purpose, we have to consider the instrumental effects
of beam size and detector noise on the number count of peaks. We here
adopt the instrumental specifications of $217$GHz channel on the future sensitive
mission {\em Planck Surveyor}; 
a Gaussian beam with full width at half maximum
 $\theta_{\rm fwhm}=5.5'$ and a pixel noise $\sigma_{\rm
pix}(\equiv s/\sqrt{t_{\rm pix}})=4.3\times10^{-6}$, where $s$ is
the detector sensitivity and $t_{\rm
pix}$ is the time spent observing each $\theta_{\rm fwhm}\times
\theta_{\rm fwhm}$ pixel.
As discussed by \cite{Takada} in detail, 
the beam smearing effect causes an incorporation of intrinsic peaks
contained within one beam, and the detector noise could make spurious
peaks in the observed CMB map. 
The noise level of Planck is sufficiently small and thus this issue will
not be so serious. For example, the peaks of height above the threshold
$\nu_{\rm s}=1$ have very significant signal to noise ratios
such as $S/N\simgt 5$. However, in the{\em MAP} case,
we have to carefully consider this detector noise effect because
the noise level per a FWHM pixel is $\sigma_{\rm pix}\sim 10^{-5}$ at
all channels of MAP and is comparable to the rms of the CMB
anisotropies field itself.
Even in this case we expect that the noise field
and the primary CMB field are statistically uncorrelated\cite{noise},
and therefore
it will be possible to perform accurate predictions of
the number count of all peaks in the observed maps including
contributions of spurious peaks mimicked by the
detector noise. This work is now in progress and will be presented
elsewhere.
Based on these considerations, for Planck case
we can approximately take into account the instrumental effects 
only by modifying the angular power spectrum as
$\tilde{C}_l=(C_l +\sigma_{\rm pix}^2\theta_{\rm
fwhm}^2)\exp[-l^2\theta_{\rm s}^2]$,
where $\theta_{\rm s}$ is expressed in terms of $\theta_{\rm fwhm}$ as
$\theta_{\rm s}=\theta_{\rm fwhm}/\sqrt{8\ln2}$.
Note that the inverse weight per solid angle,
$w^{-1}\equiv(\sigma_{\rm pix}\theta_{\rm fwhm})^2$, 
as a measure of the detector noise is the pixel-size independent
\cite{knox}. Then, the spectral parameters (\ref{eqn:spec})
can be expressed by $\sigma_{\rm n}^2(\theta_{\rm s},\sigma_{\rm pix}) 
=\int(ldl/2\pi)\tilde{C}_l l^{2{\rm n}}$.
The analytic predictions of $N_{\rm peak}$
from $\tilde{C}_l$ are indeed in good approximation with
the numerical experiments of the CMB maps \cite{HS,Takada}, where
we identified the peaks as a pixel
with higher or lower temperature fluctuations than the
surrounding pixels.
Thus it should be
noted that the number count of peaks also depends on the
instrumental effects in a general case.

For the adiabatic inflationary models,
we can accurately predict $C_l$ as a function of 
sets of cosmological parameters \cite{husugi}.
Therefore, the number count
of peaks also depends on the cosmological parameters.  
We here consider the current favored
totally flat cosmological model ($\Omega_{\rm K}\approx 0$)
as suggested by the simplest inflationary models and 
supported by the measured position of the first acoustic
peak \cite{boom,Maxima,Lange,TZ}. We then
pay a special attention to investigations of the dependence of 
$\Omega_{\rm m}$ and $h$ on the number count of peaks.
For other cosmological parameters relevant for the CMB anisotropies, 
we assume the spectral index of $n_{\rm s}=1$ for scalar
fluctuations, no tensor contribution, 
and a baryon density $\Omega_{\rm b}$ to satisfy $\Omega_{\rm
b}h^{2}=0.019$ as indicated from the big bang
nucleosynthesis measurements
\cite{tytler}. To compute the angular power spectrum for 
these cosmological models, we used the CMBFAST code 
developed by  Seljak \& Zaldarriaga \cite{cmbfast}
with the COBE normalizations \cite{Bunn}. 

Although the number count of galaxies with fixed solid angle could be 
a measure of the geometry or cosmological constant
via its dependence of the angular diameter distance \cite{GN},
this does not simply occur in the number count of 
peaks in the CMB sky as following reasons. Certainly, the comoving 
angular diameter distance at the recombination epoch $z_{\rm rec}
\approx 1100$ has a dependence of cosmological parameters as 
$D_{\rm A}\propto 1/(\Omega_{\rm m}h^{2})^{1/2}$ in the flat universe, 
and therefore except the case with  $\Omega_{\rm m}= 1$ 
we become to see a larger area on the last scattering 
surface than that in the Euclidean-like space for the fixed 
observed solid angle. However, the characteristic scale of the
temperature fluctuations to make peaks is the sound horizon scale
$r_{\rm s}$ at the last scattering and it has a similar scaling on 
the cosmological parameters such
as $r_{\rm s}\propto (\Omega_{\rm m}h^2)^{1/2}$. 
Therefore, this geometric effect is canceled out for the number count of 
peaks in the CMB map.

However we will find  that the number count of peaks has
an interesting dependence on the cosmological parameters as follows.
If recalling that the projection effect is already included into $C_l$, 
the cosmological dependence of the number count can be understood
from the $l$ dependence on power of $C_l$.
Since our model fixes the values of physical
baryon density $\omega_{\rm b}(\equiv\Omega_{\rm b}h^2)$
and $n_{\rm s}$, this means that the amplitude of the second acoustic
peak relative to the first peak is roughly
fixed \cite{husugi}. Hence there are two important physics
in the CMB anisotropies sensitive to the number count of peaks.
One is the detailed dependence of the physical matter density
$\omega_{\rm m}(\equiv\Omega_{m}h^2)$ and the vacuum density
$\Omega_{\Lambda}$ on the position of
the first acoustic peak for the flat universe.
The previous detailed
analysis \cite{husugi} revealed that lowering $\omega_{\rm m}$
causes the acoustic peaks to appear in the larger $l$, and
it simultaneously leads to a smaller characteristic curvature
scale $\theta_\ast$. Consequently, a decrease of
 $\omega_{\rm m}$ produces more peaks in smaller scales
and thus increases the number count in the observed CMB
sky as shown by equation (\ref{eqn:dn}). The effect of
lowering $\Omega_{\rm m}$ is partly compensated by the raising
of $\Omega_{\Lambda}$ in the flat universe.  Second is the driving
effect that comes from the decay of the gravitational potential 
in the radiation dominated epoch \cite{husugi}. 
Namely, a decrease of $\omega_{\rm m}$ leads to earlier
epoch of equality
between the matter and radiation energy densities and
then its driving effect enhances the acoustic oscillations and
leads to an increase in the power of anisotropies in $C_l$
at $l>l_{\rm eq}$ relative to the large scale anisotropies fixed
by the COBE normalization in our models. The change of height of
the acoustic peaks affects the number count mainly through the change
of parameter $\sigma_0$ in $\gamma$, and this effect also leads
to an increase in the number count of peaks
with lowering $\omega_{\rm m}$.  These effects complexly
affect the number count through the integration of
acoustic peaks in $C_l$ over $l$ space, and thus the constraints from
the number count on $\Omega_{\rm m}$ and $h$ cannot be equal to
those provided by $C_l$
measurements. 

\begin{figure}
\begin{center}
    \leavevmode\epsfxsize=7cm \epsfbox{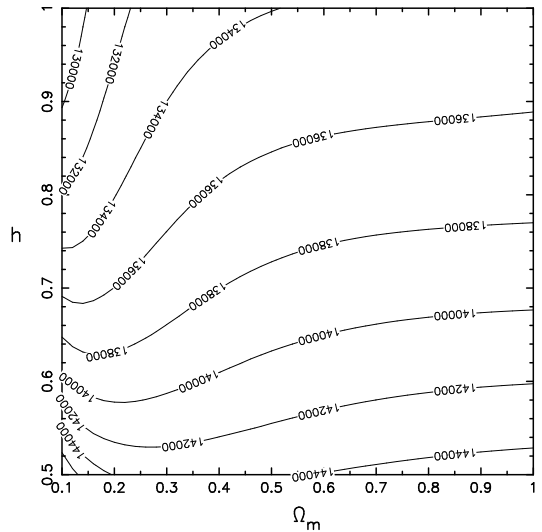}
\end{center}
\caption{The contour of
 the number count of peaks in the CMB map with the sky coverage 
 of $f_{\rm sky}=0.65$ for the expected Planck survey
 as a function of $\Omega_{\rm m}$ and $h$, where we have assumed the
 simplest flat inflationary  models with $n_s=1$ and
 $\Omega_{\rm b}h^2=0.019$. 
 We here employed the instrumental specifications of
 $\theta_{\rm fwhm}=5.5'$ and $\sigma_{\rm pix}=4.3\times 10^{-6}$. 
The contour clearly shows that lowering $\omega_{\rm m}$ roughly causes
 an increase of the number count because of the dependence of
 $\omega_{\rm m}$ on amplitude and position of the acoustic peaks of $C_l$
 in $l$-space (see text in detail). 
 }
\label{fig:peakdens}
\end{figure}

Figure \ref{fig:peakdens}
shows contours of number count of peaks as a function
of $\Omega_{\rm m}$ and $h$ for the flat $\Lambda$CDM family of
models, where we have assumed $f_{\rm sky}=0.65$
and $\nu_{\rm s}=1$ for the sky coverage and the threshold, respectively.
Although $\nu_{\rm s}=1$ is assumed for simplicity,
we have a freedom to use different measurements
of the number count with various thresholds. Furthermore, if taking
advantage of the $\nu_{\rm s}$ dependence
on the number count, we could distinguish the
contributions of spurious peaks by the detector noise from the observed
number count.  In the case of $\nu_{\rm s}=1$, the ratio of spurious
peaks to all peaks is $\sim 10\%$.
Figure \ref{fig:peakdens} shows that, although for an ideal case with
$\theta_{\rm fwhm}=\sigma_{\rm pix}=0$ the number count actually
increases with lowering $\omega_{\rm m}$ as explained, 
the dependence of $\Omega_{m}$ is reversed because of the following
reasons. The dependence  of $\Omega_{\rm m}$ on the number count
is originally very weak for the flat universe
as explained above. For example, the variation of number count
for $\Delta \Omega_{m}=0.1$ around a model with $\Omega_{\rm m}=0.3$ and
$h=0.7$ is only $\Delta N_{\rm peak}/N_{\rm peak}\approx 0.02$ in
the ideal case. The detector noise then affects more strongly
the number count in models with larger
$\Omega_{\rm m}$ for constant $h$ because those models
have lower values of $\sigma_0$. 
As a result,  the detector noise effect compensates the
variation $\Delta N_{\rm peak}$.
Most importantly, Figure \ref{fig:peakdens} shows that the
strong dependence of $h$ on $N_{\rm peak}$ still remains. 
%

\begin{figure}
\begin{center}
    \leavevmode\epsfxsize=7cm \epsfbox{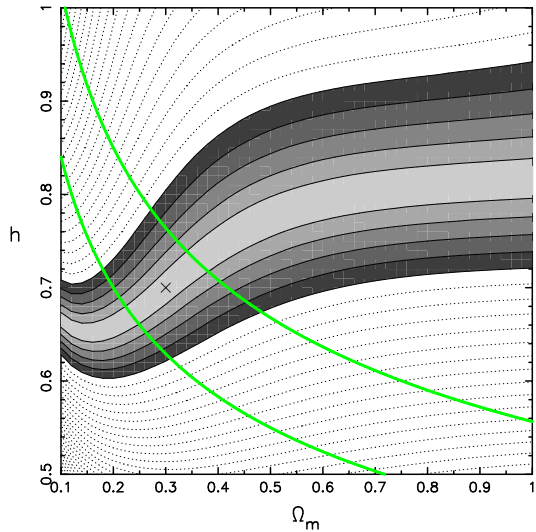}
\end{center}
\caption{The contours of the signal to noise ratio for the
 measurements of the number count of peaks for the fiducial model with
 $\Omega_{m}=0.3$ and $h=0.7$ (marked with cross)
 as same models in Figure \ref{fig:peakdens}. The fiducial model
 has $N_{\rm peak}=1.37\times10^5$. 
 The contours are stepped in units of $\Delta S/N=1$ and the
 shaded regions denote $S/N\le5$ from equation (\ref{eqn:sn}).
 The two bold solid
 lines are an arbitrarily normalized $\Omega_{\rm m}h^{3.8}$ that
 represent the dependence of constraints obtained from the
 position of the first acoustic peak in $C_l$ for the flat
 universe (Hu et al. 2000). The lines almost vertically
 cross the S/N contours.
}
\label{fig:sn}
\end{figure}

We assume that the observational errors associated
with measurements of number count of peaks with respect to a certain
threshold $\nu_{\rm s}$ can be considered as a Poisson contribution
estimated by $\sqrt{N_{\rm peak}(\nu_{\rm s})}$. We have verified using
the numerical experiments \cite{Takada} that this estimate is a good
approximation. 
We can therefore estimate the signal to noise
ratio for determinations of $\Omega_{\rm m}$ and $h $ parameters from
the number count as
\begin{equation}
\frac{S}{N}=\frac{|N^{\rm obs}_{\rm peak}(\nu_{\rm
s})-N^{\rm real}_{\rm peak}(\nu_{\rm s};\Omega_{\rm m}, h)|}
{\sqrt{N_{\rm peak}^{\rm obs}(\nu_{\rm s})}},
\label{eqn:sn}
\end{equation}
where the subscripts ``obs'' and ``real'' denote the observed value 
and theoretical prediction of the number count, respectively. 
Based on this consideration, Figure {\ref{fig:sn}} shows contours of
constant signal to noise ratio $S/N$ for determinations of $\Omega_{\rm m}$ and
$h$ for the fiducial model with $\Omega_{\rm m}=0.3$ and $h=0.7$ marked 
with cross. The fiducial model then has $N_{\rm peak}=1.37\times 10^5$
and the contours or equivalently constraints on $\Omega_{m}$ and
 $h$ approximately scale as $\Omega_{\rm }h^{-4.9}$
around the point of the fiducial model.
Recently, Hu et al. (2000) shows that the combined data of $C_l$
from BOOMERanG and MAXIMA can place the constraints on
these parameters, which approximately scale as $\Omega_{\rm
m}h^{3.8}$, from the measured position of
fist acoustic peak around $l\approx200$ under the assumption of
a flat universe (Figure 6 in their paper). The dependence
of $\Omega_{\rm m} h^{3.8}$ is also shown by two bold solid lines in
Figure \ref{fig:sn} and it clearly demonstrates that the lines
almost vertically cross the $S/N$ contours of the number count.  
Therefore, combining both constraints allows us to
accurately determine $\Omega_{\rm m}$ and $h$ or break the cosmic degeneracy
by the CMB measurements alone. 

In this {\it Letter}, we propose a new potentially useful method
that the number count of peaks in the CMB map can probe the
cosmological parameters relatively independently of those
provided by the measurements of $C_l$. This is because the peak
statistics contains the integrated information of $C_l$ over $l$ space,
although the number count can be derived only from $C_l$
based on the Gaussian theory. Thus, if we adopt the Gaussian assumption
for the primordial fluctuations such as the inflationary scenarios
suggest, we could extract an additional information on the
underlying cosmology. As an example, we here showed that the number
count of peaks can place complementary constraints on $\Omega_{\rm m}$-$h$
plane and an interesting possibility that we can determine
$\Omega_{\rm m}$ by combining the
constraints from $C_l$. This result therefore indicates that
we can break the cosmic degeneracy
by using the CMB data alone without invoking the other astronomical
observations.

Undoubtedly, secondary anisotropies and foregrounds can mimic the peaks
in the observed CMB maps. Although the most important sources are the thermal
Sunyaev-Zel'dovich effect, this effect can be removed by either
observing  at $217$GHz or using advantage of its specific spectral
property. We also have to carefully investigate the effect of Galactic
foregrounds and extragalactic point sources \cite{foregrounds}
to make reliable predictions in our method, 
but this must be done for any measurements
of statistical properties of CMB temperature map. 

We thank Eiichiro Komatsu for valuable and critical comments, which
considerably improved this manuscript. 
We are grateful to U.~Seljak and M.~Zaldarriaga for making 
available their CMBFAST code. 
M.T. acknowledges a support from a JSPS fellowship.   


\begin{thebibliography}{99}
\bibitem{noise}However, the detector noise tends to make more spurious
peaks on smooth structures nearer to intrinsic peaks in the CMB sky.
\end{thebibliography}
\end{document}